\documentclass[aps,preprint,superscriptaddress,showpacs,floatfix]{revtex4-1}
\usepackage{amsfonts,amssymb,latexsym,xspace,epsfig,graphicx,color}
\usepackage{amsmath,enumerate,stmaryrd,xy,stackrel}

\usepackage[caption=false]{subfig}
\usepackage{ulem}
\include{srctex}
\usepackage{xfrac}

\usepackage{ulem}
\newcommand{\ecal}{\mathcal{E}}
\newcommand{\kcal}{\mathcal{K}}

\begin{document}
\title{Stationary Real Solutions of the Nonlinear Schr\"odinger Equation on a Ring with a Defect}

\author{Axel P\'erez-Obiol}
\affiliation{Laboratory of Physics, Kochi University of Technology, Tosa Yamada, Kochi 782-8502, Japan}
\author{Taksu Cheon}
\affiliation{Laboratory of Physics, Kochi University of Technology, Tosa Yamada, Kochi 782-8502, Japan}

\begin{abstract}
We analyze the 1D  cubic nonlinear stationary Schr\"odinger equation on a ring 
 with a defect for both focusing and defocusing nonlinearity. All possible $\delta$ and $\delta'$ boundary conditions are considered at the defect, computing for each of them the real eigenfunctions, written as Jacobi elliptic functions, and eigenvalues for the ground state and first few excited energy levels. All six independent Jacobi elliptic functions are found to be solutions of some boundary condition. We also provide a way to map all eigenfunctions satisfying $\delta$/$\delta'$ conditions to any other general boundary condition or point-like potential.
\end{abstract}

\date{\today}

\maketitle

\section{Introduction}

The 1D nonlinear Schr\"odinger equation (NLSE) has been studied in
various topologies, including the infinite line, the half-line,
the ring, the box, star graphs, the tadpole diagram,
and H and Y networks \cite{zakharov72,fokas89,carr00,tokuno08,gattobigio12,hung11,adami11,cacciapuoti15, mastr18}. These studies are mainly motivated
by the success of the NLSE in describing a wide range of nonlinear phenomena.
Most relevantly, light propagating in nonlinear fibers and Bose-Einstein
condensates (BECs) in the quasi-1D limit. In BECs,
various topologies can be realized through different confining traps,
such as of toroidal shape, and barriers can be added with laser beams
or weak links \cite{gupta05,morizot06,ryu07,engels07,ramanathan11}.
Moreover, with Feshbach resonances \cite{cornish00}, the nonlinearity coupling
can be tuned through a wide range of negative (attractive) and positive
(repulsive) values.

In this work we study the full spectrum of real stationary solutions
for a ring with a general 
 point-like defect.
The 1D NLSE can be solved through the inverse scattering transform,
or by directly integrating the equation and writing the solutions
as Jacobi elliptic functions. These two methods were respectively first applied
in the infinite line \cite{zakharov72} and in the ring and the box \cite{carr00},
and thereafter in many other works.
The case of a ring with a 
 point-like defect was first
explored theoretically in \cite{nakamura17}.
In particular, part of the eigenfunctions and spectra
were found for a delta-type defect using various forms of Jacobi functions.
This calculation was done for only part of the parameter space
describing the delta boundary conditions, and some gaps were found in the spectrum.
Using a simpler set of Jacobi functions, with elliptic modulus
always constrained between 0 and 1, we extend this work to the complete
set of delta conditions, finding that both eigenfunctions and spectra vary
continuously in all parameter space.
We then compute the case of $\delta'$ conditions, and finally generalize to
any real boundary condition, finding thus all the real stationary solutions
for a ring with a general point-like potential for the ground and first excited states.

Any point-like defect might be described
as a certain type of boundary condition .
In turn, a general boundary condition can be divided into four
subsets, two of them so called $\delta$ and $\delta'$.
We show that the four sets of normalized eigenfunctions satisfying
any of the four types of boundary conditions, or in fact any general condition,
are actually the same
set of functions. By computing all possible eigenfunctions satisfying
the $\delta$ conditions, we are therefore computing all possible solutions
of the boundary problem. This is explicitly demonstrated by mapping
the $\delta$ case into the $\delta'$ one.
We thus put emphasis in the $\delta$ and $\delta'$ conditions,
showing the energy spectra for the ground state and first excited
energy levels for 
both conditions and the relation between them.

The rest of the paper is organized as follows:
 in section ~\ref{sec:nlse}, we present the boundary problem and specify
the equations to be solved. First, we integrate the NLSE and then determine the
normalization and the boundary conditions and the relations between them. 
The end of the section is devoted to the scaling of the NLSE.
In section \ref{sec:limits} we analyze
the linear and no-defect limits.
The eigenvalues and eigenfunctions for the nonlinear case are presented in section~\ref{sec:results}.
In the final section we discuss our results.

\section{
Nonlinear Schr\"odinger equation with a defect}
\label{sec:nlse}

Let us consider the stationary NLSE on a 1 dimensional ring of length $L$ and with a defect
at $x=0$. Assuming a real wave function $\phi$, nonlinear parameter $g$,
and energy $E$, it reads, 
\begin{align}
\label{eq:NLSE}
-\phi''+g\,\phi^3=E \phi.
\end{align}
Integrating once, one gets $-{\phi'}^2+\frac{g}{2}\phi^4=E\,\phi^2-2c$,
where $c$ is an integration constant. Note that $c$ must be real 
for $\phi$ to be real too.
Completing the square of the non-derivative term and integrating again, the equation becomes,
\begin{align}
\label{eq:int1}
\int_0^{\tilde\phi(x)}\frac{d\phi}{\sqrt{(\phi^2-k_{-}^2)(\phi^2-k_{+}^2)}}
=&\sqrt{\frac{g}{2}}(x-x_0),
\end{align}
where $x_0$ is another integration constant that fixes $\tilde\phi(x_0)=0$,
and $k_\pm=\sqrt{\frac{E}{g}\pm\sqrt{\frac{E^2}{g^2}-\frac{4c}{g}}}$.
Changing variables with $\phi=k_{-}\sin(\theta)$, one obtains the
elliptic integral of first kind,
\begin{align}
\int_0^{\arcsin(\tilde\phi(x)/k_{-})}\frac{d\theta}{\sqrt{1-\sfrac{k_{-}^2}{k_{+}^2}\,\sin(\theta)^2}}
=\sqrt{\frac{g}{2}}k_{+}(x-x_0),
\end{align}
which allows us to define the NLSE solution in terms of the Jacobi sine sn, 
\begin{align}
\tilde\phi(x)=k_{-}\text{sn}\left(\sqrt{\sfrac{g}{2}}\, k_{+}(x-x_0),\sfrac{k_{-}^2}{k_{+}^2}\right).
\end{align}
Redefining the integration constant $c$ to $m=\sfrac{k_{-}^2}{k_{+}^2}$,
we have,
\begin{align}
\label{eq:phi}
\phi_{sn}(x)=\frac{1}{\sqrt{g}}\sqrt{\frac{2\,E\,m}{1+m}}\,{\rm sn}\left(\sqrt{\frac{E}{1+m}}(x-x_0),\, m\right),
\end{align}
where all the square roots are real if $g>0$, $E>0$, and $0<m<1$.
Note that the integration parameter $m$ plays the role of the elliptic modulus,
which defines the shape of the Jacobi function, and the nonlinear parameter
$g$ normalizes the function.
Since $c$ is real, $m$ is constrained to be either real (when $E^2\ge 4\,c\,g$) or complex with modulus $|m|=1$ (for $E^2<4\,c\,g$).
Moreover, for real $m$, Jacobi transformations allow to constrain it to $0<m<1$.

By making different changes of variables in Eq.~(\ref{eq:int1}) one can find that all the 12 Jacobi functions solve the NLSE. Out of these 12, only 6 have different shapes (not related through scaling and shifting).
These functions can be defined in a general form as
$\phi_{pq}(x)=\frac{1}{\sqrt{g}}\,\sqrt{\alpha}\,{\rm pq}\left(\sqrt{\beta}(x-x_0),\, m\right)$,
where pq $=$ sn, cn, dn, ns, cs, and ds defines the type of Jacobi function
and with coefficients
$\alpha$ and $\beta$ such that $\sqrt{\alpha/g}$
and $\sqrt{\beta}$ are always real.
Table~\ref{tab:phis} shows the explicit expressions of $\alpha$ and
$\beta$ for each pq together with the domains of $g$ and $E$ that each function solves. All of them have $0<m<1$ except for sn, which
also allows for complex modulus with $|m|=1$. ns may also have
complex modulus and $|m|=1$, but in this case it only differs from sn
through a shifting and an overall sign.
In particular, $\phi_{cn}$ and $\phi_{dn}$ solve the attractive case ($g<0$) and
the others the repulsive one ($g>0$).
\begin{table}	
\begin{center}
\begin{tabular}[c]{c|c|c|c}
\ pq  \ &  $\alpha$  &  $\beta$ & domain
\\ [0.3ex]  
\hline
sn  &  \ $\frac{2\,E\,m}{1+m}$  \ & \ $\frac{E}{1+m}$ \ & \ $g>0$, $E>0$
\\ [0.9ex]  
\hline
cn  &  \ $\frac{2\,E\,m}{2m-1}$  \ & \ $\frac{E}{1-2m}$  \ &  \ $g<0$, $E$ real
\\  [0.9ex]  
\hline
dn  &  \ $\frac{2\,E}{2-m}$  \ & \ $\frac{-E}{2-m}$  \ & \ $g<0$, $E<0$
\\  [0.9ex]  
\hline
ns  &  \ $\frac{2\,E}{1+m}$  \ & \ $\frac{E}{1+m}$  \ & \ $g>0$, $E>0$
\\ [0.9ex] 
\hline
cs  &  \ $\frac{-2\,E}{2-m}$  \ & \ $\frac{-E}{2-m}$  \ & \ $g>0$, $E<0$
\\ [0.9ex] 
\hline
ds  &  \ $\frac{2\,E}{1-2m}$  \ & \ $\frac{E}{1-2m}$  \ & \ $g>0$, $E$ real
\end{tabular}
\end{center}
\caption{Coefficients $\alpha$ and $\beta$ defining the six independent solutions
of the NLSE and the domains $g$ and $E$ they solve.}
\label{tab:phis}
\end{table}

\begin{figure}[t]
\subfloat{\includegraphics[width=.4\textwidth]{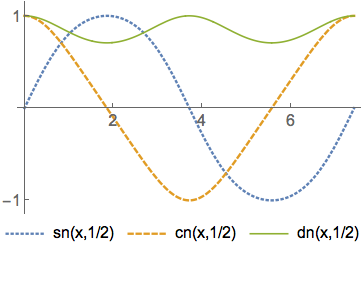}}
~~
\subfloat{\includegraphics[width=.4\textwidth]{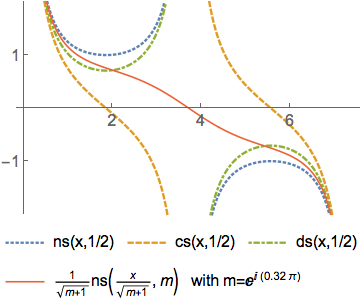}}
\caption{(Color online). Convergent (left) and divergent (right) Jacobi elliptic functions. Values of $m=\frac12,\,e^{i\,0.32\pi}$ taken for illustration purposes.}
\label{fig:jac}
\end{figure}

$\phi_{cn}$, $\phi_{sn}$, and $\phi_{dn}$ are convergent while
$\phi_{cs}$, $\phi_{ns}$, and $\phi_{ds}$ are divergent as long
as $0<m<1$ (see Fig.~\ref{fig:jac}).
In the cases of box and periodic boundary conditions, the period or semiperiod
of the wave function is fixed to be exactly $L$, and divergent Jacobi
functions are not allowed in order to avoid singularities. On the contrary, for general boundary conditions,
the period might be small enough to contain only the convergent parts
and all the six functions are possible solutions.

\subsection{Normalization and boundary conditions}

The normalization is inversely proportional to the nonlinear parameter $g$,
since $\phi^2\propto 1/g$, and as can be
seen by a rescaling of $\phi$ in Eq.~(\ref{eq:NLSE}).
We fix the normalization to 1,
\begin{align}
\label{eq:norm}
\int_0^L dx\, \phi(x)^2=1,
\end{align}
with $L=2\pi$, and leave $g$ as a parameter.

Excluding disconnected cases, a general point boundary condition
can be parametrized as
\begin{align}
\label{eq:genbc}
\left({
\begin{array}{c}
\phi_L
\\
\phi_L'
\end{array}
}\right)
=&
e^{i\theta}
\left({
\begin{array}{cc}
a & b
\\
c & d
\end{array}
}\right)
\left({
\begin{array}{c}
\phi_0
\\
\phi_0'
\end{array}
}\right),
\end{align}
with $\theta\in [0,\pi)$, $a$, $b$, $c$, $d\in\Re$ and $a\,d-b\,c=1$ \cite{albeverio98}. $\phi_0$ and $\phi_L$ are defined as the limits $\phi_{0}=\lim_{\epsilon\to0}\phi(\epsilon)$,
$\phi_{L}=\lim_{\epsilon\to0}\phi(L-\epsilon)$, and the same
for the derivatives.
This general condition ensures flux conservation and 
can be explicitly constructed as short-range limits of three neighboring Dirac $\delta$-functions \cite{CS98}.

For real wave functions ($\theta=0$),
this family of boundary conditions
 can be further subdivided into the four subsets
$a=0$, $b=0$, $c=0$, and $d=0$, here labeled as $A$, $B$, $C$, and $D$.
Each subset is itself a two-parameter family of boundary conditions.
For fixed non-linearity, the solutions of the NLSE depend on
three parameters, $E$, $m$, and $x_0$.
One is fixed by the normalization, and the two left can be
fixed by any of the four two-parameter subsets of boundary conditions.

A real eigenfunction that satisfies any general boundary condition
will also satisfy any of the four subsets (for certain
values of the two corresponding parameters). Moreover, this function
 is a solution for a unique pair of parameters in each subset.
To show this we impose the constraints of Eq.~(\ref{eq:genbc})
with $\theta=0$
for the general condition and for the four subsets at the same time and get,
\begin{align}
\label{eq:mappinggen}
\left({
\begin{array}{c}
\phi_L
\\
\phi_L'
\end{array}
}\right)=
&\left({
\begin{array}{cc}
a & b
\\
c  & d
\end{array}
}\right)
\left({
\begin{array}{c}
\phi_0
\\
\phi_0'
\end{array}
}\right)
\\
\label{ee1}
%\nonumber
%%%%%%%%%%%%%%%%%%%%%%%%%%
=
&\left({
\begin{array}{cc}
0 & b_{A} 
\\
c_{A}  & d_{A} 
\end{array}
}\right)
\left({
\begin{array}{c}
\phi_0
\\
\phi_0'
\end{array}
}\right)
%%%%%%%%%%%%%%%%%%%%%%%%%%
%\nonumber
\\
\label{ee2}
%\nonumber
=&\left({
\begin{array}{cc}
a_{B} & 0
\\
c_{B}  & d_{B} 
\end{array}
}\right)
\left({
\begin{array}{c}
\phi_0
\\
\phi_0'
\end{array}
}\right)
%%%%%%%%%%%%%%%%%%%%%%%%%%
\\
\label{ee3}
%\nonumber
=&
\left({
\begin{array}{cc}
a_{C} & b_{C} 
\\
0 & d_{C} 
\end{array}
}\right)
\left({
\begin{array}{c}
\phi_0
\\
\phi_0'
\end{array}
}\right)
\\
\label{ee4}
%\nonumber
=&
\left({
\begin{array}{cc}
a_{D} & b_{D} 
\\
c_{D}  & 0
\end{array}
}\right)
\left({
\begin{array}{c}
\phi_0
\\
\phi_0'
\end{array}
}\right).
%%%%%%%%%%%%%%%%%%%%%%%%%%
\end{align}
To find the values of the two parameters for one the types of boundary conditions,
given a function $\phi(x)$ that solves another type or the general one,
we solve the corresponding (consistent) system of linear equations in
Eqs.~(\ref{eq:mappinggen})-(\ref{ee4}), and obtain
\begin{align}
\label{eq:bcinv}
b_A&
= -\frac{1}{c_A} 
= b+a\,\mu,
&d_A=&d+c\mu+\frac{\mu}{b+a\mu};
\\
\nonumber
a_B&
= \frac{1}{d_B}
= a+\frac{b}{\mu},
&c_B=&c+\frac{d}{\mu}-\frac{1}{b+a\mu};
\\
\nonumber
a_C&
= \frac{1}{d_C}
=\frac{1}{d+c\mu},
&b_C=&b+a\mu-\frac{\mu}{d+c\mu};
\\\nonumber
a_D&
=a+\frac{b}{\mu}+\frac{1}{d+c\mu},
&b_D
=& -\frac{1}{c_D}
=-\frac{\mu}{d+c\mu};
\end{align}
with $\mu\equiv\frac{\phi_0}{\phi_0'}$.
This means that the four types of conditions
(\ref{ee1})-(\ref{ee4}) 
have, as solutions, the same group of eigenfunctions. 
Therefore, if one of the sets is completely solved, the others can be directly mapped using Eq.~(\ref{eq:bcinv}).
The same applies to the general condition (\ref{eq:mappinggen}),
which can also be mapped to any of the four conditions with a single zero entry.

In this paper we focus on 
(\ref{ee2}) and (\ref{ee3}), the so called $\delta$ and $\delta'$
boundary conditions, respectively, which we reparametrize as
\begin{align}
\label{eq:dbc}
t\, \phi(0)-\phi(L)&=0,
\\\nonumber
\phi'(0)-t\, \phi'(L)&=v\, \phi(0);
\end{align}
and
\begin{align}
\label{eq:dpbc}
t'\, \phi(0)-\phi(L)&=v'\,\phi'(L),
\\\nonumber
\phi'(0)-t'\,\phi'(L)&=0.
\end{align}
A solution $\phi(x)$ satisfying Eq.~(\ref{eq:dbc})
will also solve Eq.~(\ref{eq:dpbc}) for $t'$, $v'$, such that
\begin{align}
\label{eq:mapping}
t'=t+v\frac{\phi(0)}{\phi'(L)},~~~
v'=v\left(\frac{\phi(0)}{\phi'(L)}\right)^2.
\end{align}

Within these two families,
important special cases are
Dirichlet ($\phi_0=\phi_L=0$, $v\to\pm\infty$),
Neumann ($\phi_0'=\phi_L'=0$, $v'\to\pm\infty$),
Dirac delta of strength $v$ ($\phi_0=\phi_L$, $\phi_0'-\phi_L'=v\phi_0$, $t=1$),
first derivative of the Dirac delta ($t'\phi_0=\phi_L$, $\phi_0'=t'\phi_L'$, $v'=0$),
and the continuity condition ($\phi_0=\phi_L$, $\phi_0'=\phi_L'$,
$t=1$, $v=0$ or $t'=1$, $v'=0$).
To illustrate the relation between the boundary condition and the point like
defect one may integrate the nonlinear Schr\"odinger equation with the
corresponding potential. For example, in the continuous ($\phi(0)=\phi(L)$) case with a Dirac delta,
integrating around around $x=0$,
\begin{align}
\label{eq:NLSE}
\int_{-\epsilon}^{\epsilon} dx\left(
-\phi''(x)+g\,\phi(x)^3+v\delta(x)\phi(x)-E \phi(x)\right)=0,
\end{align}
and taking the limit $\epsilon\to0$, one obtains the $\delta$ boundary
condition,
$\phi'(0)-\phi'(L)=v\,\phi(0)$.

\subsection{Scaling of the NLSE}
The NLSE is invariant under $\tilde{x}=\lambda x$,
$\tilde{g}=\lambda^2 g$, and $\tilde{E}=\lambda^2\,E$. This 
can be shown by scaling $x\to \tilde{x}=\lambda\, x$ in Eq.~(\ref{eq:NLSE}),
\begin{align}
\label{eq:scaling}
-\phi''(\tilde{x})+g\,\lambda^2\phi(\tilde{x})^3=\lambda^2E\,\phi(\tilde{x}),
\end{align}
with $\phi''(\tilde{x})=\frac{\partial^2 \phi(\tilde{x})}{\partial \tilde{x}^2}$.
Therefore, for any $\phi(x)$ satisfying the NLSE,
a function $\phi(\lambda\,x)$ is also a solution
with $\tilde{g}=\lambda^2\,g$ and $\tilde{E}=\lambda^2\,E$.
$\phi(\lambda\,x)$ does not satisfy in general the
boundary conditions and normalization, $N=\int_0^L dx\,\phi(\lambda\,x)^2\neq1$.
However, the renormalized function $\frac{1}{\sqrt{N}}\phi(\lambda\,x)$ satisfies
 Eq.~(\ref{eq:scaling}) with $\tilde{g}\to N \tilde{g}$,
\begin{align}
-\frac{\phi''(\tilde{x})}{\sqrt{N}}+g\,\lambda^2\,N \left(\frac{\phi(\tilde{x})}{\sqrt{N}}\right)^3=E\,\frac{\phi(\tilde{x})}{\sqrt{N}}.
\end{align}
The $\delta$ and $\delta'$ boundary conditions in Eqs.~(\ref{eq:dbc}) and~(\ref{eq:dpbc}) are invariant
under scaling as long as $v$ and $v'$ are reparametrized to $\lambda\,v$ and $\frac{1}{\lambda}\,v'$, respectively,
and the scaling amounts to a shifting of an integer number ($n$) of periods ($T$) in the Jacobi functions at $x=L$, so that
$\phi(\lambda L)=\phi(L)$, $\phi'(\lambda L)=\phi'(L)$ (and trivially $\phi(\lambda 0)=\phi(0)$, $\phi'(\lambda 0)=\phi'(0)$).
This constrains the scaling factor
to $\lambda=\lambda_n=1+\frac{T\,n}{\sqrt{\beta}L}$, where
$\beta$ depends on the Jacobi function as in Table~\ref{tab:phis}.

For every solution, we then have a new set of normalized solutions that also satisfy the boundary
conditions with $g\to N\,\lambda_n^2\,g$, $E\to\lambda_n^2\,E$, and $v\to\lambda_n\,v$ for $\delta$ and
$v'\to\frac{v'}{\lambda_n}$ for $\delta'$.

\section{Linear and no defect limits}
\label{sec:limits}

\subsection{Linear limit}
\label{sec:lineal}

In the linear limit, $g=0$, the NLSE becomes $-\phi''=E\,\phi$, and a general solution
can be written as $\phi(x)=A\,\sin(k (x-x_0))$, with $A$ an integration constant
fixed by normalization and $k\equiv\sqrt{E}$. Using $\delta$ boundary conditions
given by Eqs.~(\ref{eq:dbc}), $x_0$ reads
\begin{align}
\label{eq:x0l}
x_0=-\frac{1}{k}\arctan\left(\frac{\sin(L\,k)}{t-\cos(L\,k)}\right),
\end{align}
where $k$ is fixed by
\begin{align}
\label{eq:deltag0}
t=\frac{1\pm\sqrt{\sin(L\,k)^2-\sfrac{v}{k}\sin(L\,k)\cos(L\,k)}}{\cos(L\,k)}.
\end{align}
Similarly, for $\delta'$, $x_0'$ and $t'$ read
\begin{align}
x_0'=-\frac{1}{k}\arctan\left(\frac{k-t'\cos(L\,k)}{\sin(L\,k)}\right),
\end{align}
\begin{align}
\label{eq:deltapg0}
t'=\frac{1\pm\sqrt{\sin(L\,k)^2+v'\,k\sin(L\,k)\cos(L\,k)}}{\cos(L\,k)}.
\end{align}
Using Eqs.~(\ref{eq:deltag0}) and~(\ref{eq:deltapg0}) we find $t(E,v)$
and $t'(E,v')$, and then obtain numerically $E(t,v)$ and $E(t',v')$.
Note that these two equations are related
through $v'=-\frac{v}{k^2}$, and therefore $E(t',v')=E(t,-\frac{v}{k^2})$ can also be computed
directly from $E(t,v)$ (and vice versa).
In terms of the mapping between $\delta$ and $\delta'$ of Eq.~(\ref{eq:mapping}),
we have, using Eq.~(\ref{eq:x0l}),
$\frac{\phi(0)}{\phi(L)}=\frac{\sin(k L)}{k(t \cos(k L)-1)}$ and
\begin{align}
t'=&\,t+\frac{v \sin(k L)}{k(t \cos(k L)-1)},
\\\nonumber
v'=&\,\frac{v \sin(k L)^2}{k^2(t \cos(k L)-1)^2}.
\end{align}

\begin{figure}[t]
\subfloat{\includegraphics[width=0.4\textwidth]{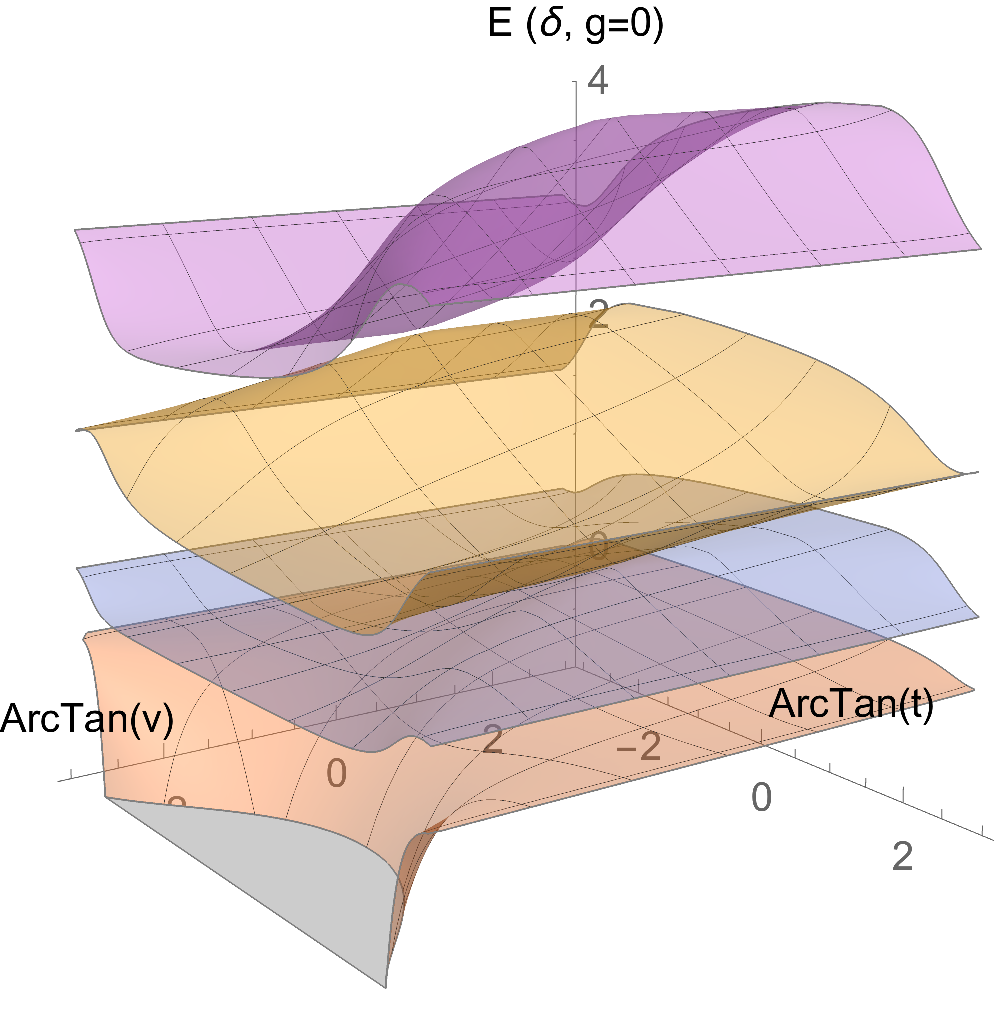}}
~~
\subfloat{\includegraphics[width=0.4\textwidth]{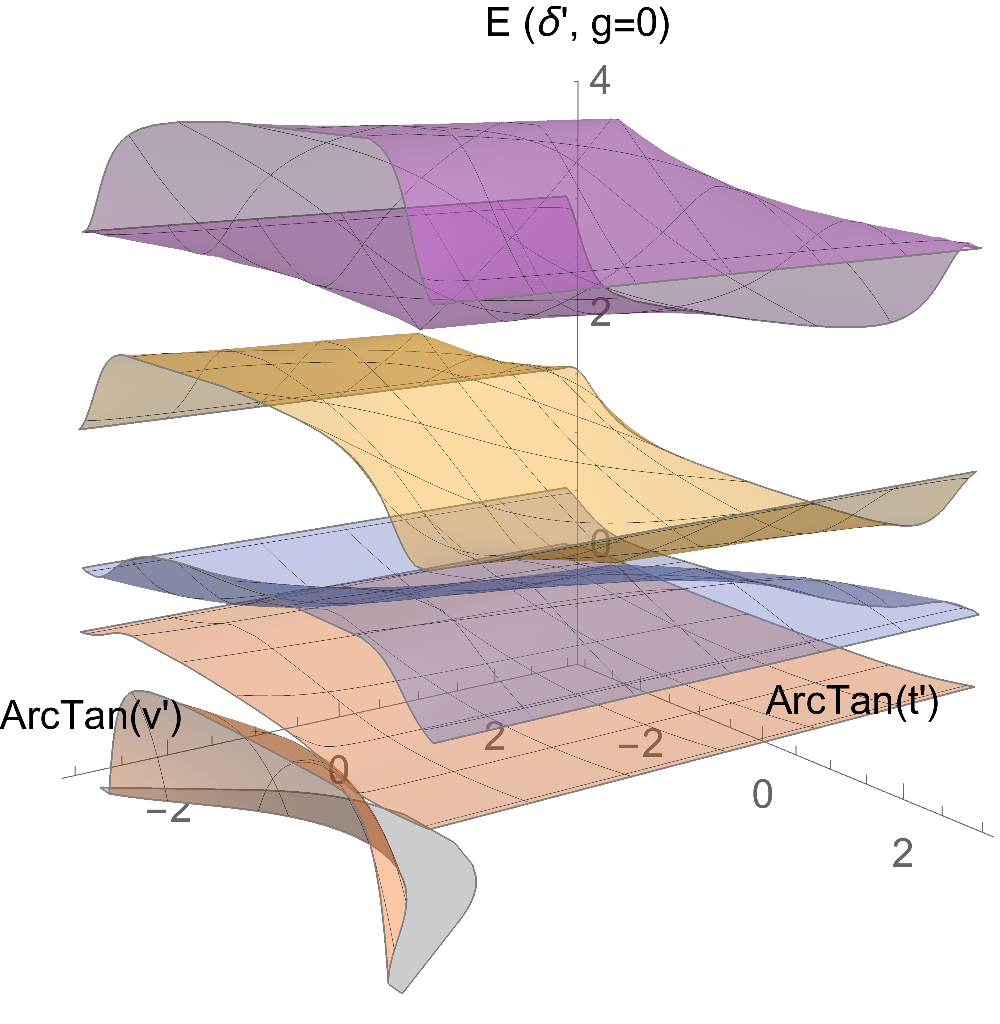}}
\caption{(Color online). Energy spectra for $g=0$ (linear case)
and $\delta$ (left) and $\delta'$ (right) boundary conditions.
The energy levels are colored
according to how the levels in $\delta$ are mapped onto $\delta'$.}
\label{fig:deltag0}
\end{figure}

Fig.~\ref{fig:deltag0} shows the energy spectra
$E(t,v)$ and $E(t',v')$ for the first four energy levels and $g=0$.
For both boundary conditions a foam-like energy structure is found,
each energy level being a continuous surface in all parameter space $(t,v)$ and $(t',v')$.
The energy levels are only connected through a point in the lines $t=\pm 1$, $v=0$ and $t'=\pm 1$, $v'=0$.
As $v\to -\infty$, the $\delta$ boundary conditions represent a $\delta$ type
potential with depth going to negative infinity, and as expected,
the lowest energy level diverges. In the $\delta'$ case we have
that the energy diverges in the limit $v'=\frac{-v}{k^2}\to0$.

\subsection{No defect limit}
\label{sec:nodef}

\begin{figure}
\includegraphics[width=.6\textwidth]{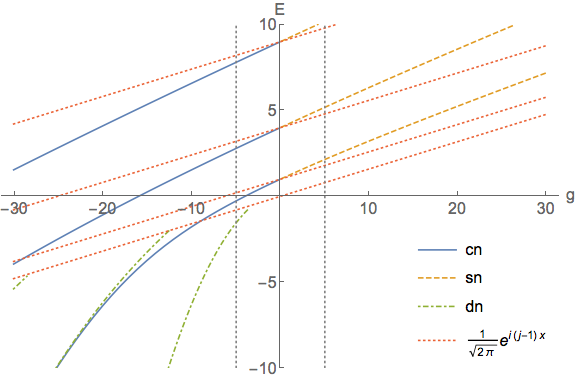}
\caption{(Color online). Energy spectra as a function of nonlinearity $g$,
for $t=1$, $v=0$ (no defect case), and real wave functions and plane waves.
The gray vertical lines indicate the energy spectra for $g=\pm 5$.}
\label{fig:nodefect}
\end{figure}

The NLSE on a ring without defect has already been analyzed thoroughly in \cite{carr00}.
Here we review it for completeness and as a basis for the non-periodic
boundary conditions. Imposing continuity on the eigenfunctions fixes
their period to $L$, and $x_0$ becomes redundant and can be taken $x_0=0$.
There are then only three possible independent solutions, cn, sn, and dn,
\begin{align}
\phi_{sn}^0(x)=&
\frac{\sqrt{m}}{\sqrt{L}\sqrt{1-\frac{\ecal(m)}{\kcal(m)}}}
\text{sn}\left(\frac{2j\kcal(m)}{\pi}\,x,m\right),
\\
\phi_{cn}^0(x)=&
\frac{\sqrt{m}}{\sqrt{L}\sqrt{m-1+\frac{\ecal(m)}{\kcal(m)}}}
\text{cn}\left(\frac{2j\kcal(m)}{\pi}\,x,m\right),
\\
\phi_{dn}^0(x)=&
\frac{\sqrt{\kcal(m)}}{\sqrt{L}\sqrt{\ecal(m)}}
\text{dn}\left(\frac{2j\kcal(m)}{\pi}\,x,m\right),
\end{align}
which solve the NLSE for $m$ and $E$ that satisfy, respectively,
\begin{align}
\pi^2 g=&8 j^2 L\, \kcal(m)\left[(\kcal(m)-\ecal(m)\right],
\label{eq:msn}\\\label{eq:mcn}
\pi^2 g=&8 j^2 L\, \kcal(m)\left[(1-m)\kcal(m)-\ecal(m)\right],
\\\label{eq:mdn}
\pi^2 g=&-2 j^2 L\, \kcal(m) \ecal(m);
\end{align}
and,
\begin{align}
E=&\frac{4j^2}{\pi^2}(1+m)\kcal(m)^2,
\label{eq:esn}\\\label{eq:ecn}
E=&\frac{4j^2}{\pi^2}(1-2m)\kcal(m)^2,
\\\label{eq:edn}
E=&-\frac{j^2}{\pi^2}(2-m)\kcal(m)^2;
\end{align}
with $j$ a positive integer and $\kcal(m)$ and $\ecal(m)$ the complete
elliptic integrals of first and second kind.
By solving for $m$ in Eqs.~(\ref{eq:msn})-(\ref{eq:mdn}) and inserting  it
into Eqs.~(\ref{eq:esn})-(\ref{eq:edn}) we find the energy spectra as
a function of $g$ (see Fig.~\ref{fig:nodefect}).
Note that sn appear as solutions for $g>0$,
cn for $g<0$, and dn exist only for $g\leq-\pi j^2$.
The latter is due to Eq.~(\ref{eq:mdn}) and that $\kcal(m)\ecal(m)\geq \frac{\pi^2}{4}$
for all real $m$.
In the point where dn functions emerge, $g=-\pi j^2$, the energies are $E=-\frac{j^2}{2}$,
and they coincide with the spectrum from the trivial solution
$\phi(x)=\frac{1}{\sqrt{2\pi}}$, $E=\frac{g}{2\pi}$.
This constant function can also be interpreted as the first of the plane wave
solutions, $\frac{1}{\sqrt{2\pi}}e^{i\,(j-1)\,x}$, with spectra
$E=g/(2\pi)+(j-1)^2$, also plotted in Fig.~\ref{fig:nodefect}.

\section{Results}
\label{sec:results}
\subsection{Energy spectra}
\label{sec:energies}

In order to compute the energy spectra $E(t,v)$ for $g\neq0$ and fix the parameters $m$ and $x_0$
in the eigenfunctions we need to solve Eqs.~({\ref{eq:norm}) and
(\ref{eq:dbc}).
For this we use the Newton and quasi-Newton methods integrated in Wolfram Mathematica,
which require an initial guess close enough to the solution.
Given an initial solution, one can use it as a guess for a point 
that is close enough in parameter space.
Since the three parameters $E$, $m$ and $x_0$ are found to be continuous
for each energy level,
one can swipe all $(t,v)$ in a systematic way to compute $E(t,v)$.
Fig.~\ref{fig:spectrad} shows the four first energy levels
for $g=-5$ and $g=5$.
The structure of the spectrum is similar to the linear case,
with the exception of the bubble stemming from the bottom level for $g=-5$.
Computing $E(t,v)$ for various $g$ between
$g=0$ and $g=-5$, we find that this bubble emerges at $g=-\pi$.
This is expected from Section~\ref{sec:nodef}, since for periodic boundary
conditions dn functions appear at $g=-\pi j^2$, and $t=1$, $v=0$
are the minima and maxima of the energy levels. Therefore,
other bubbles are expected to appear at $g=-4\pi,-9\pi,-16\pi$, etc.
\begin{figure}
\subfloat{\includegraphics[width=.4\textwidth]{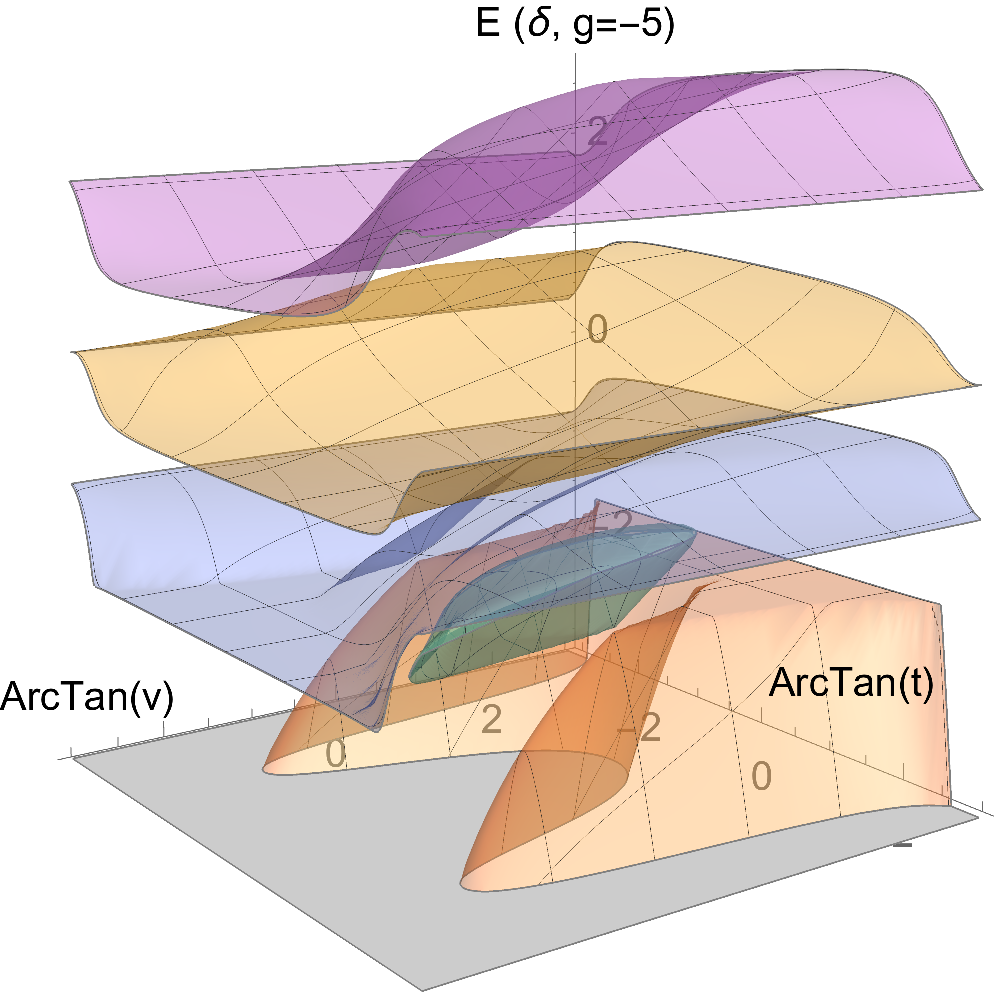}}
~~
\subfloat{\includegraphics[width=.4\textwidth]{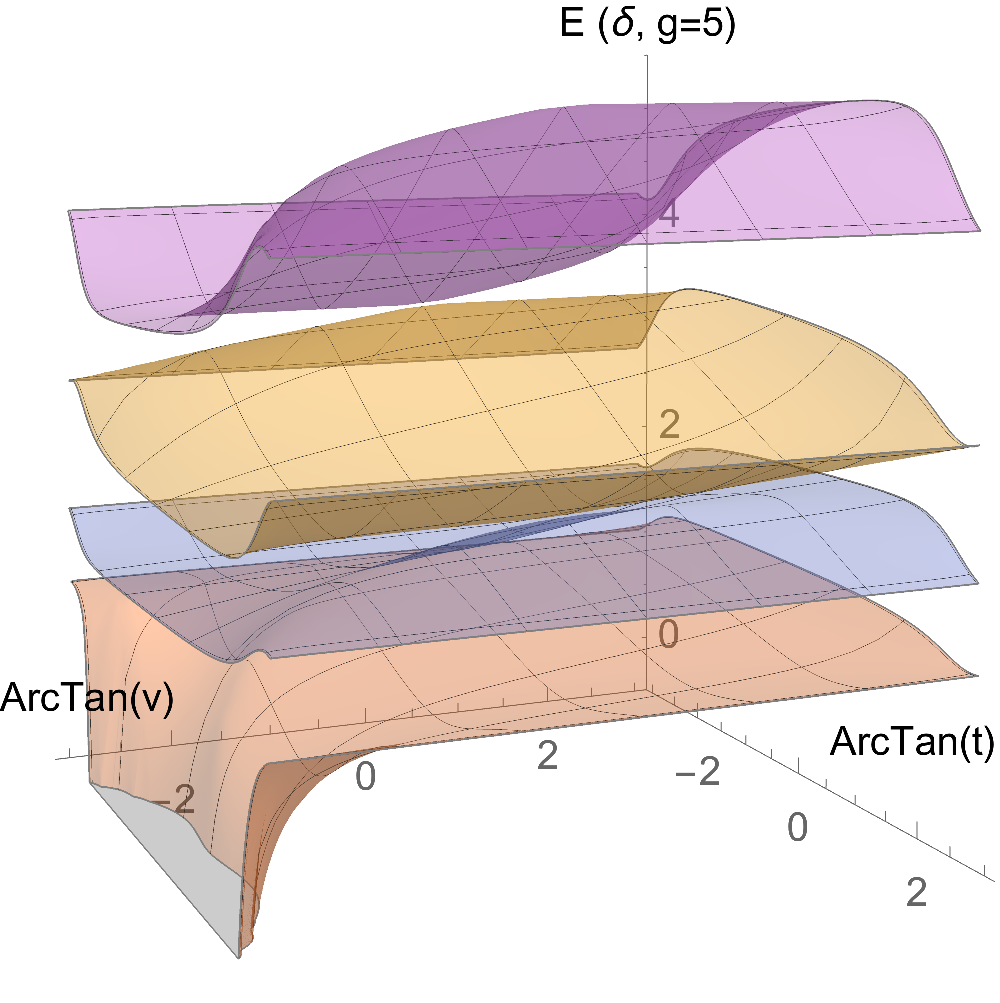}}
\caption{(Color online). Energy spectra for $\delta$ boundary conditions and for $g=-5$ (left) and $g=5$ (right).}
\label{fig:spectrad}
\end{figure}

The $\delta'$ case can in principle be computed in the same way.
However, the energy spectrum is much more complex, presenting bubbles in both
the attractive and repulsive cases, which avoid a smooth swiping
of the parameter space. One can instead take the spectra and eigenfunctions
already computed for $\delta$, and map them to $E(t',v')$
using Eq.~(\ref{eq:mapping}). We find that this mapping is complete,
the whole spectrum in $\delta$ filling all the energy levels in $\delta'$
and vice versa. The corresponding spectra is plotted
in Fig.~\ref{fig:spectradp}.
\begin{figure}[t]
\centering
\subfloat{
\includegraphics[width=.3\textwidth]{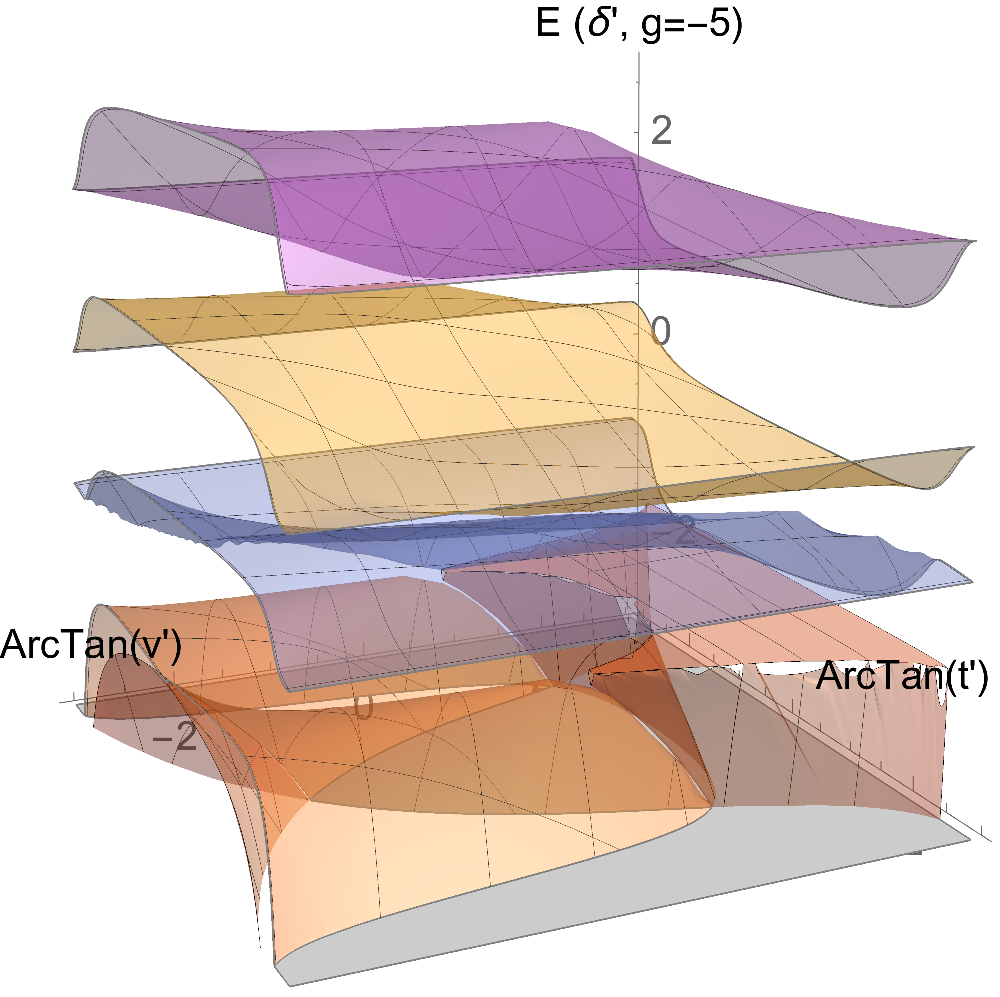}
~
\includegraphics[width=.3\textwidth]{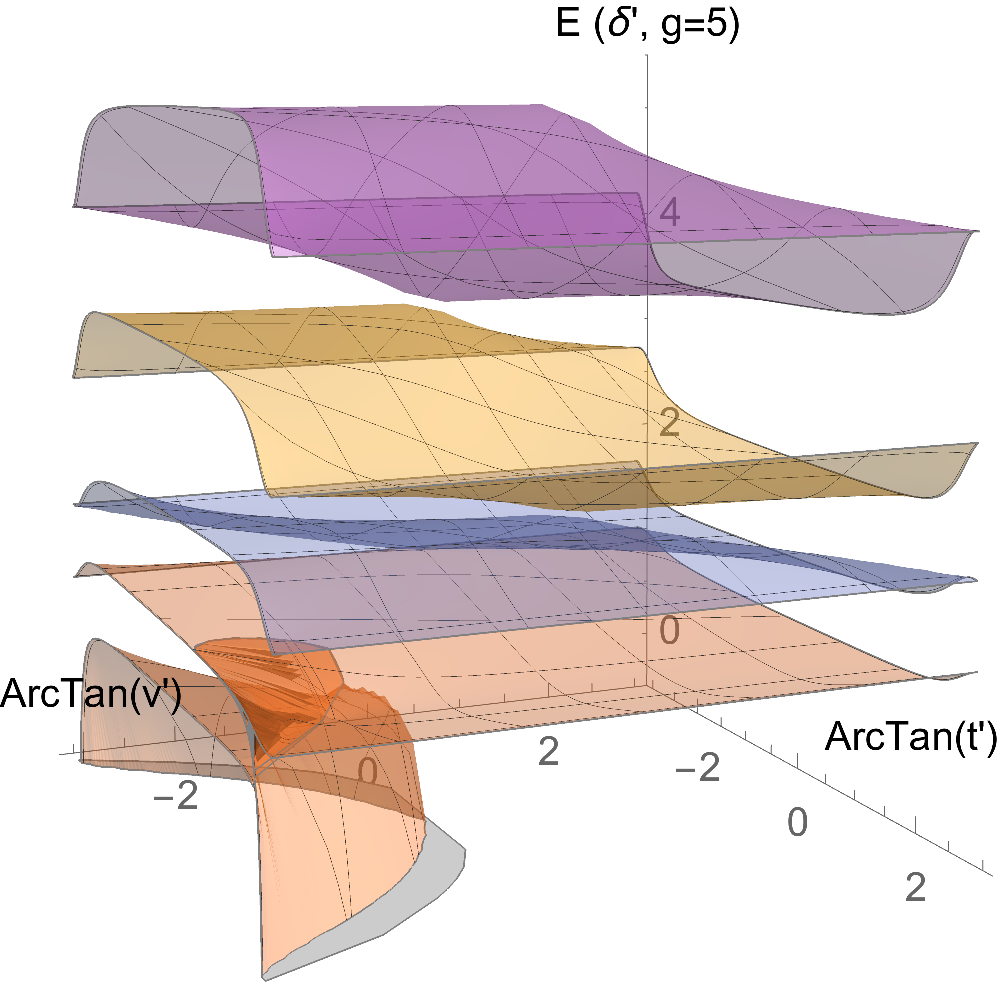}
~
\includegraphics[width=.3\textwidth]{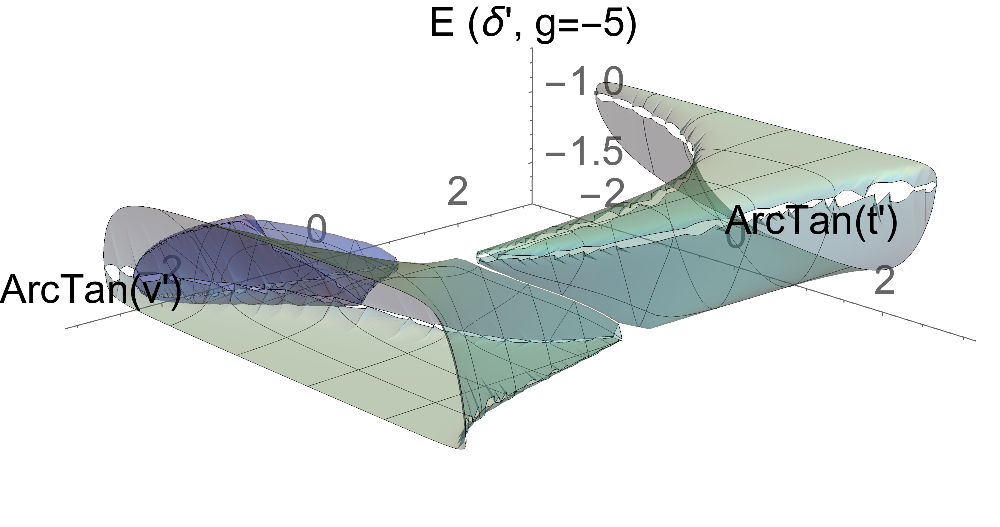}}
\caption{(Color online). Energy levels depending on $t'$ and $v'$ for $g=-5$ (left) and $g=5$ (middle).
Surfaces are colored as the corresponding energy levels for $\delta$
boundary conditions. The bubbles in the right figure are plotted separately
for better visualization, and they fit right below the lowest flat energy level in the plot in the very left.}
\label{fig:spectradp}
\end{figure}
The positive part of the energy spectrum is alike to the one
with $\delta$ boundary conditions: continuous energy levels which
increase with $g$. However, the negative part of the spectrum
presents a much richer bubble-like structure, specially for $g=-5$.
Note that $\delta$ and $\delta'$ boundary conditions include
the periodic ($t=t'=1$, $v=v'=0$) and antiperiodic
($t=t'=-1$, $v=v'=0$) ones, where the spectra in both
cases coincides. These points are also the minima and maxima
of the flat energy levels and where they overlap.

\subsection{Eigenfunctions}
\label{sec:eigenfunctions}

All six independent Jacobi functions
appear as solutions of the NLSE.
Each one satisfies a different set of boundary conditions
described by a region in the parameter space ($t$,$v$).
Due to the mapping between boundary conditions, each region
in ($t$,$v$)
has a corresponding one in ($t'$,$v'$).
The boundaries between these regions in the space ($t$, $v$)
correspond to the limiting cases $m=0$ and $m=1$.
For these values of the elliptic modulus, the Jacobi functions
at each side of the boundary become the same trigonometric
function, for example,
\begin{align}
\nonumber
\phi_{ns}(x,0)=&
\phi_{ds}(x,0)=
\phi_{ds}(x,1)=
\phi_{cs}(x,1)=
\\=&
\sqrt{2\,E/g} \, {\rm csc}\left(\sqrt{E}(x-x_0)\right).
\end{align}
By solving Eqs.~(\ref{eq:norm}) and~(\ref{eq:dbc}) with $m=0,1$
we find the boundaries $t(v)$ separating the various regions for
the ground state in $\delta$,
as shown in Fig.~\ref{fig:region}.
For $g=-5$, the distribution is simple, cn and dn with $0<m<1$ being grossly divided
between $t<0$ and $t>0$, respectively.
The case of $g=5$ has much more structure, with sn, ns, cs, ds with
$0<m<1$ and ns with complex modulus $|m|=1$ solving different regions
(see Fig.~\ref{fig:region}).
The bubble appearing between the ground state and the first excited
flat energy level in the attractive case corresponds entirely to dn type eigenfunctions.
All the upper flat energy levels have eigenfunctions of type cn
when $g=-5$, and sn for $g=5$.

Excited energy levels are related to the number of crests and troughs of the wave functions,
with each higher level containing one more of them. In these cases, the eigenfunctions periods are smaller
than $L$ and the solutions are constrained to the convergent ones, sn, cn and dn, in accordance
to the distribution of eigenfunctions stated above. Moreover, dn and cs are not allowed
at excited levels with positive energies, since their domains are restricted to $E<0$
(see Table~\ref{tab:phis}).

The solutions may also be characterized through their parity. Under a parity transformation,
$x\to L-x$, the NLSE and the normalization stay invariant, while the boundary conditions stay
the same only for $t=\pm1$, $t'=\pm1$. In this case, the ground state and the bubble for $\delta$ conditions
have an even eigenfunction for all $v$. For $v<0$, odd excited states are even
functions, while the even ones are odd, and vice versa for $v>0$.
Within this subset of eigenfunctions with definite parity, an orthogonal relation can be defined
between even and odd solutions.
Both the parity and the number of crests and troughs of the eigenfunctions are illustrated
in Fig.~\ref{fig:eigenfunctions} for $t=1$, $g=\pm5$, and $v=\pm1$.
\begin{figure}[t]
\centering
\subfloat{\includegraphics[width=.4\textwidth]{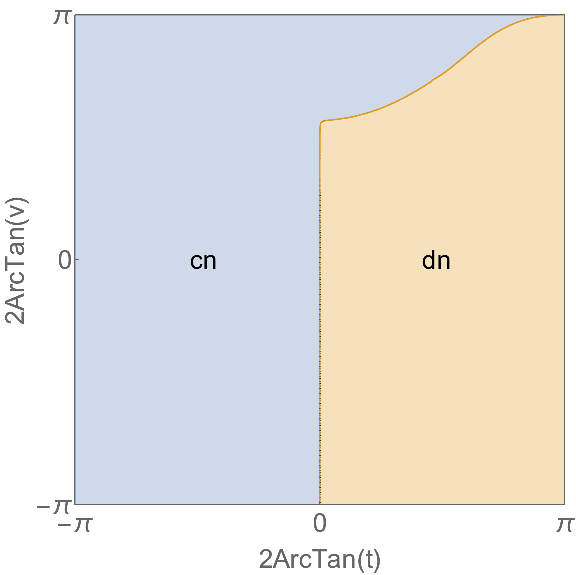}}
~~
\subfloat{\includegraphics[width=.4\textwidth]{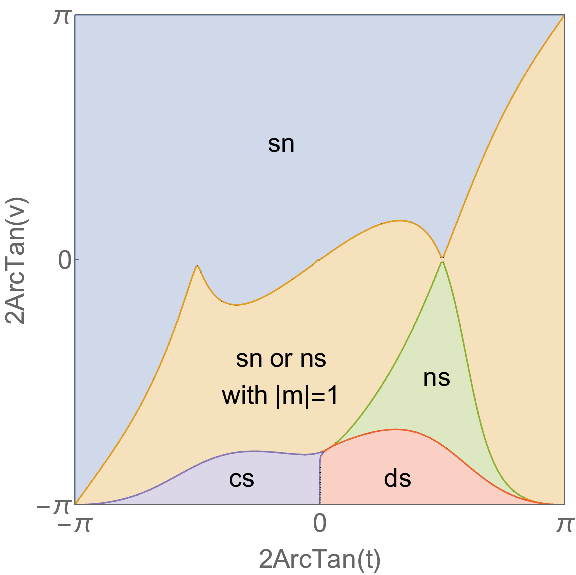}}
\caption{(Color online). Regions of the parameter space ($t$, $v$) 
according to the type of Jacobi function that satisfies them in the ground state for $g=-5$ (left) and $g=5$ (right). The upper flat levels are satisfied by cn for $g=-5$ and sn for $g=5$. The bubble in the attractive case corresponds entirely to dn functions.}
\label{fig:region}
\end{figure}

\begin{figure}[t]
\centering
\subfloat[$g=-5$, $v=-1$.]{\includegraphics[width=0.24\textwidth]{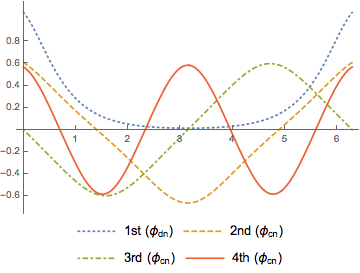}}
\subfloat[$g=-5$, $v=1$.]{\includegraphics[width=0.24\textwidth]{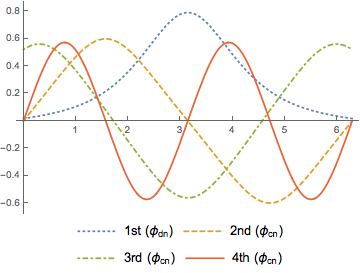}}
\subfloat[$g=5$, $v=-1$.]{\includegraphics[width=0.24\textwidth]{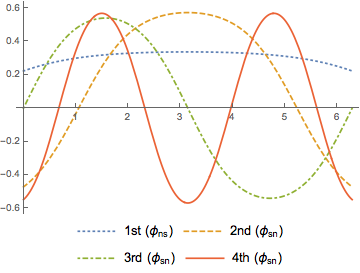}}
\subfloat[$g=5$, $v=1$.]{\includegraphics[width=0.24\textwidth]{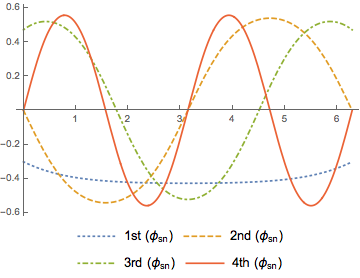}}
\caption{(Color online). Eigenfunctions with well defined parity for the first four (flat) energy levels at $t=1$, $g=\pm5$, $v=\pm1$.}
\label{fig:eigenfunctions}
\end{figure}

\section{Conclusions}

This article tackles the boundary problem of the time-independent NLSE on a 1D ring with real wave functions normalized to one,
nonlinear parameters $g=-5$ and $g=5$,
and focusing on $\delta$ and $\delta'$ connection conditions.
The solutions of the NLSE are written in the form of the six independent Jacobi elliptic functions
 sn, cn, dn, ns, cs, and ds. Their elliptic modulus is constrained to be either real with $0<m<1$, or
 complex with $|m|=1$.
We find that all six functions are necessary in order to solve all possible $\delta$ and $\delta'$ boundary conditions.
In particular, cn and dn solve the attractive case, $g=-5$, and sn, cs, ns, and ds the repulsive one, $g=5$.
Only sn or ns type of solutions have also complex elliptic modulus.
These possible 7 types of functions are mapped into the parameter space defining all $\delta$ connections, ($t$, $v$),
for $g=-5$ and $g=5$, and each energy level.

The continuity of the energy spectrum and integration constants through
all parameter spaces ($t$, $v$) and ($t'$, $v'$) makes the problem
solvable for all possible boundary $\delta$ and $\delta'$ conditions in the ground and first excited states.
The function $E(t,v)$ presents, for both $g=-5$ and $g=5$, a series of flat energy levels, with the bottom level diverging as $v\to-\infty$, interpretable in terms of
a $\delta$ interaction with depth going to $-\infty$. For $g=-5$ a bubble
appears on top of the bottom level, contributing to two more energy levels
of nodeless (dn type) wave functions.
This bubble and similar ones
emerge at $g=-4\pi,-9\pi,-16\pi$, etc., and can be understood as the
generalization of the dn spectra in the no defect limit from Fig.~\ref{fig:nodefect}.
The energy spectra for $\delta'$ shows a much richer topology,
specially for $g=-5$.
In this case, apart from a series of excited flat energy levels, 
a few bubbles, each adjacent to each other, appear for $E<0$.
Due to the correspondence with the $\delta$ case, we know that two
of these bubbles also appear at $g=-\pi$.
A fixed range of energies might then present a different amount
of eigenvalues depending on the type of defect and nonlinearity.
In general, 
all the energy surfaces are continuous and present no holes.
When increasing $g$, the levels rise and the spacing between them diminishes.
The attractive case presents a richer structure
in the eigenvalue spectra, while the repulsive one has a more complex
distribution of eigenfunction types.

In all cases, the flat energy levels coincide with the immediate lower and upper ones
in the case of periodic (continuum limit) or antiperiodic conditions, $t=t'=\pm 1$, $v=v'=0$. These two particular cases correspond to two points
in our parameter space, and their solutions consist in only three types
of functions: cn and dn for the attractive nonlinearity, and sn for the repulsive one \cite{carr00}. Notably, in the repulsive case,
a defect allows for six new types of eigenfunctions (the five divergent ones
plus sn with $|m|=1$).

The topologies of the energy spectra $E(t,v)$ and $E(t',v')$ might
look very different, but the set of energies and eigenfunctions are found to be exactly the same for
$\delta$ and $\delta'$ and fixed $g$. They are, however, when mapped from one parameter space to the other, redistributed in a quite convoluted way.
We emphasize that this mapping has been very convenient in the computation of
$E(t',v')$. Moreover, it might also prove useful
to solve the NLSE with the other two sets of point boundary conditions,
and in general, any point-like connection.

\section*{Acknowledgments}
This work was supported by the Japan Ministry of Education, Culture, Sports, Science and Technology under the Grant number 15K05216.

\end{document}